\documentclass[preprint]{aastex}

\shorttitle{New high proper motion stars at low galactic latitudes}
\shortauthors{Lepine et al.}

\begin{document}

\title{New High Proper Motion Stars from the Digitized Sky Survey.
I. Northern Stars with $0.5<\mu<2.0\arcsec yr^{-1}$ at Low Galactic
Latitudes.\altaffilmark{1,2,3}}

\author{S\'ebastien L\'epine\altaffilmark{1,2} and Michael M. Shara}
\affil{Department of Astrophysics, Division of Physical Sciences,
American Museum of Natural History, Central Park West at 79th Street,
New York, NY 10024, USA}

\and

\author{R. Michael Rich\altaffilmark{1}}
\affil{Department of Physics and Astronomy, University of California
at Los Angeles, Los Angeles, CA 90095, USA}

\altaffiltext{1}{This paper is respectfully dedicated to the memory of
the late Barry M. Lasker, former director of the Catalogs and Survey
Branch of the Space Telescope Science Institute, whose vision was
instrumental in the creation of the Digitized Sky Survey and Guide Star
Catalogs.}

\altaffiltext{2}{Based on data mining of the Digitized Sky Survey,
developed and operated by the Catalogs and Surveys Branch of the Space
Telescope Science Institute, Baltimore, USA.}

\altaffiltext{3}{Based on data mining of the Guide Star Catalog-II, a
joint project of the Space Telescope Science Institute and the
Osservatorio Astronomico di Torino.}

\begin{abstract}
We have conducted a systematic search for high proper motion stars in
the Digitized Sky Survey, in the area of the sky north of -2.8 degrees
in declination and within 25 degrees of the galactic plane. Using the
SUPERBLINK software, a powerful automated blink comparator developed
by us, we have identified 601 stars in the magnitude range $9<r<20$
with proper motions in the range $0.5<\mu<2.0$ arcsec yr$^{-1}$ which 
have not been observed with the HIPPARCOS satellite. Among those, we
recovered 460 stars previously listed in Luyten's proper motion
catalogs (LHS, NLTT), and discovered 141 new high proper motion stars,
ranging in magnitude from R=13.0 to R=19.8. Only 9 stars from the
Luyten catalogs that were not observed by HIPPARCOS could not be
recovered with SUPERBLINK: 3 had proper motions larger than the search
limit of the code ($\mu>2.0\arcsec yr^{-1}$), and 5 were missed because
they were either too bright for SUPERBLINK to handle or they are in
the immediate proximity of very bright stars. Only one of Luyten's
stars (LHS1657) could not be recovered at all, even by visual
inspection of the POSS plates, and is now suspected to be bogus. The
very high success rate in the recovery by SUPERBLINK of faint Luyten
stars suggests that our new survey of high proper motion stars is at
least 99\% complete for stars with proper motions $0.5<\mu<2.0$ arcsec
yr$^{-1}$ down to R=19. This paper includes a list of positions,
proper motions, magnitudes, and finder charts for all the new high
proper motion stars.
\end{abstract}



\section{Introduction}

The study of stars with large proper motions forms the basis of our
knowledge of the stellar content and dynamical structure of the
Galaxy. It is largely through surveys of high proper motion (HPM)
stars that the census of stars in the solar vicinity has been
established. Since the proper motion of a star is inversely
proportional to its distance from the observer, a high proper motion
is a strong selection criterion for proximity. The power of proper
motion as a selection tool is well illustrated when we consider that
among the billions of stars brighter than about 20th magnitude that
fill the sky, those with proper motions $\mu$ larger than half of a
second of arc per year ($\mu>0.5\arcsec yr^{-1}$) are numbered in the
thousands.

The stellar proper motion is also proportional to the velocity of the
star, projected on the plane of the sky. This is the main drawback for
samples of nearby stars derived from HPM surveys: stars which are
moving toward or away from the Sun are systematically missed. Another
effect is that samples of HPM stars are always over-represented with
stars having large transverse velocities. Essentially, the high velocity
stars are sampled over a larger volume than the low velocity
stars. While this may bias statistical studies of HPM samples, it is
actually extremely useful for finding old disk and halo stars, which
are rare in the solar neighborhood, and crucial to our understanding of
the structure and stellar content of the Galaxy as a whole.

Of course, direct parallax measurements are ultimately the best method
to measure the distances to the stars. However, even the most complete
survey to date, carried out by the Hipparcos astrometric satellite,
is limited to the brightest hundred thousand stars in the sky
(brighter than about 12th magnitude, see the {\it Hipparcos and Tycho
catalogs} Perryman 1997). Next-generation astrometric mission are
being planned (the GAIA mission, see Lindegren \& Perryman 1996; the
DIVA mission, see R\"oser 1999) which will eventually obtain
parallaxes for hundreds of millions more stars. But the identification
of nearby stars in the 12th-20th magnitude range still depends largely
on surveys of high proper motion stars.

To date, the most extensive catalogs of high proper motion stars
remain the two catalogs by W. H. Luyten, published over 20 years
ago. One, the {\it LHS catalogue: a catalogue of stars with proper
motions exceeding 0.5$\arcsec$ annually} (Luyten 1979) lists 3602
objects with estimated proper motions $\mu\geq0.500\arcsec$ yr$^{-1}$
and 867 other stars with estimated proper motions
$0.235\leq\mu<0.500\arcsec$ yr$^{-1}$. The {\it LHS catalogue} was
compiled from previous lists of known HPM stars and from a list of new
HPM stars detected in the 1950's Palomar Sky Survey by hand-blink or
with an automated blink-machine. The complete catalog is available at the
ViZier service of the {\it Centre de Donn\'ees Astronomiques de
Strasbourg} (http://vizier.u-strasbg.fr/) under catalog number
I/87B. The {\it New Luyten Catalogue of stars with proper motions
larger than two tenths of an arcsecond} (NLTT, Luyten 1979) lists
58845 HPM stars, the majority of which have estimated proper motions
$\mu\geq0.18\arcsec$ yr$^{-1}$. The NLTT catalogue is essentially an
extension of the LHS catalogs to stars with smaller proper motions,
and is also available at the ViZier service under catalog number
I/98A.

Both the LHS and NLTT catalogs were first published in a printed
version based on a typewriter copy of the catalog, and both are thus
subject to occasional typos and misprints, only a few of which are
obvious. The entire LHS catalog has been recently investigated by
Bakos, Sahu, \& N\'emeth (2002), who searched for each and every one of
the LHS stars in the Digitized Sky Survey. While they were able to
recover the majority of the LHS stars, they found substantial errors
in the coordinates quoted by Luyten. A small number of LHS stars could
not even be recovered at all, raising serious doubts as to the
reliability of the Luyten catalogs. Furthermore, the LHS and NLTT are
also clearly incomplete in some parts of the sky, especially in the
southern hemisphere and at low galactic latitudes. The incompleteness
of the Luyten catalogs, and the large inaccuracies it sometimes
contains, make it a less than completely reliable tool in an epoch where
data mining of large electronic databases is becoming an important
aspect of astronomical research.

While other surveys of high proper motion stars have been conducted
since the publication of the LHS and NLTT catalogs, these surveys have
been more limited in magnitude range or in survey area. The
most notable is the survey conducted with the Hipparcos satellite
(Hipparcos and Tycho catalog, ESA 1997; Tycho-2 catalog, Hog {\it et
al.} 2000) which covers the whole sky but is complete only for
relatively bright ($V\lesssim10$) stars. Deeper large proper motion
surveys include the UCAC survey conducted by the US. Naval Observatory,
which reaches $V\approx16$ and covers the whole sky south of
$\delta=+24$, and the SuperCOSMOS Sky Survey (Hambly {\it et al.}
2001) which so far covers 5000 square degrees in the southern galactic
cap (15\% of the sky) down to $R\approx19$. Other deep surveys
covering smaller areas include the search for very high proper motion
stars in 1400 square degrees of second epoch Palomar Sky Survey plates
by Monet {\it et al.} (2000), the Calar-ESO proper motion survey (Ruiz
{\it et al.} 2001), and the southern survey by Wroblewski \& Costa
(2001). Other proper motion surveys have been conducted using fields
monitored by gravitational microlensing experiments: the EROS 2 proper
motion survey (EROS collaboration 1999), and a search of the MACHO data
archive (Alcock {\it et al.} 2000). 

For the past few years, we have been conducting our own systematic,
automated survey for high proper motion stars using the Digitized Sky
Survey (DSS). We are searching for high proper motion stars in the DSS
with SUPERBLINK, a powerful software developed by one of us
(SL). SUPERBLINK is an extremely efficient, fully automated,
blink-comparator which can be used to identify variable and moving
objects in any pair of FITS images of a field observed at two
different epochs. We have been feeding DSS images to our SUPERBLINK
software and assembling a new catalog of HPM stars, complete with
``blinkable'' double-epoch finder charts. This first paper of a series
presents some of the first results of our efforts: an updated census
of HPM stars with $\mu>0.5\arcsec$ yr$^{-1}$ down to a magnitude
R=20.0 north of -2.8$^{\circ}$ in declination and at low galactic
latitudes ($|b|<25^{\circ}$). We report the recovery of 640 Luyten
stars and the discovery of 141 new HPM stars within this area. A
second series of papers (L\'epine et al. 2002, in preparation) will
present the results of our parallel campaign of spectroscopic
observations of those new high proper motion stars.

In \S2, we describe the data retrieval, image handling, candidate
identification, and photometry. \S3 and \S4 contrast our results with
the Tycho and Luyten catalogs. Finder charts for all the new high
proper motion stars are printed in the appendix.


\section{A new survey of high proper motion stars using the Digitized
Sky Survey}

\subsection{Data retrieval}

We are retrieving Digitized Sky Survey (DSS) images from the Space
Telescope Science Institute archive server (archive.stsci.edu) using
Linux scripts, with the kind permission of the Catalogs and Surveys Branch
at the STScI. Our first epoch images the scanned Palomar Sky Survey
(POSS-I, Abell 1959) red (xx103aE + plexi) images which
have been obtained circa 1950 and have been scanned with a resolution
of 1.7$\arcsec$ pixel$^{-1}$. The second epoch images, the scanned
Second Palomar Sky Survey (POSS-II, Reid {\it et al.} 1991) red (IIIaf
+ RG 610) images are generally of higher quality, and have been
scanned with a resolution of 1.00 to 1.02 $\arcsec$ pixel$^{-1}$. We
adopted the following strategy: the POSS-I image is used as a
reference frame onto which we attempt to match the POSS-II image, to
generate two image that look as similar as possible and that can be
easily subtracted to reveal the differences between the two epochs.

The first epoch survey covers the entire sky north of
$\delta=-2.7^{\circ}$ in a grid of $6.5^{\circ}\times6.5^{\circ}$
plates with a small overlap region at the sides of the plates (643
plates in all). The POSS-II plates cover the same region with plates
of similar sizes, but with a larger overlap between neighboring
plates, and thus with a larger number of plates to cover the northern
sky. As a result, there is no direct correspondence between the POSS-I
and POSS-II plates, and the same region of the sky can appear in very
different locations on the first and second epoch plates. This means
that larger scale distortions in the plates can be significantly
different in the POSS-I and POSS-II. Since we are using a
differential technique to identify high proper motion stars, the
correspondence between the two fields must be extremely tight. 

For these reasons, we are working on small subfields of the Digitized
Sky Survey representing areas on the POSS-I and POSS-I plates no larger
than $17\arcmin\times17\arcmin$ in size. We arrived at this figure
after intensive experimentation with DSS subfields. We found that
superposition of POSS-I and POSS-I fields of this size is reasonably
accurate and straightforward, and can be easily handled by the
software. We have mapped the whole sky north of $\delta=-2.7^{\circ}$
into 642,900 subfields, allowing for a significant overlap between the
fields.

\subsection{Image superposition}

The SUPERBLINK software uses optimization techniques to align the two
images and to degrade and normalize the second epoch images so that they
match the aspect and quality of the earlier epoch image. The code then
simply subtracts the two images and identifies significant features in
the residuals. In order for the residual image to contain significant
information, the two images must be aligned very precisely. The process
is illustrated in Figure 1.

To align the images, the code proceeds in a series of steps. First the
image with the highest angular resolution per pixel element is
remapped onto a grid that matches the resolution of the other image.
In our data, it is the POSS-II image (with resolution of 1.00, 1.01,
or 1.02$\arcsec$ pixel$^{-1}$) that is remapped into a grid that
matches the POSS-I image (1.70$\arcsec$ pixel$^{-1}$). This results
in a small degradation in the quality of the POSS-II image, but this
is of little consequence since the quality of the POSS-II image is
always better to start with.

The second step is to normalize the two images to match the intensity
levels. For each image, pixels are separated into {\it sky} pixels and
{\it star} pixels. The mean pixel value and standard deviations are
calculated; any pixel that is above 3 times the mean standard
deviation is flagged as a {\it star} pixel and removed from the
set. The process is repeated recursively until all the pixels are
flagged as either {\it star} or {\it sky}. The sky pixels are then
fitted to a plane and the whole image is renormalized so that the mean
value of the sky is set to 1. The total flux from star pixels is then
calculated, and each image is renormalized so that the fluxes above
the sky level are equal.

In the third step, the software attempts to minimize the total
residuals between the two images by shifting and rotating the higher
quality (POSS-II) image. We use a multidimensional minimization
algorithm to determine the shift in X and Y and the rotation $\Theta$
that matches the two images best. We retrieve the POSS-I and POSS-II
images based on the coordinates at the center of the images, so we
expect the two images to be already aligned in X and Y. In effect we
find that the centers of the two images are not always perfectly
aligned, and a typical shift of a fraction of a pixel (rarely
exceeding 1 pixel) is usually required. On the other hand, the two
images can be at an angle because they may come from different parts
of a POSS-I and POSS-II plate (e.g. upper left corner of a POSS-I
plate with lower right corner of a POSS-II plate) which means that the
individual images can be oriented differently, with the north pointing
in different directions. We thus sometimes require large values for
$\Theta$, up to 40 degrees in high declination fields.

\subsection{Identification of moving objects}

High proper motion stars are identified as pairs of residual
features. The SUPERBLINK software contains a search algorithm
which attempts to match any significant residual feature with another
one up to a given distance. While the task appears to be relatively
simple, it is actually quite complex because one wants to avoid
the systematic identification of any differences between the plates as a
high proper motion star. The POSS-I and POSS-II images are scans of
photographic plates, and they contain numerous plate
defects. Furthermore, the images are filled with large numbers of
variable features, from variable stars (and nebulae!), to asteroid
tracks, diffractions spikes and instrumental reflections from bright
stars. The POSS-I and POSS-II plates have also been obtained in
slightly different bandpasses (the POSS-I plate used an xx103aE
emulsion with a plexiglas filter while the POSS-II used a IIIaF
emulsion with an RG610 filter) which means that some objects are
expected to have slightly different plate magnitudes. The POSS-II
plate also goes a few magnitudes deeper, and some of the fainter stars
do not record on the POSS-I plates. For all these reasons, specific
tests have been designed within the code to eliminate most unwanted
features while retaining a set of the most likely high proper motion
star candidates.

At the same time, the code remains a general tool for identifying
moving objects of any shape and size (i.e. no attempt is made to fit
the variable features with an axisymmetric stellar profile). The
reason for this is that high proper motion stars tend to be so
close to the Sun that long-period binary systems are often
resolved. Because the PSF on the POSS images is so broad (exposure
times $>30$ minutes), companions may look like a moving elliptical
or pear-shaped object, which can easily be confused with plate defects
or short asteroid tracks.

The general procedure to test whether a feature in the residual image
is consistent with a high proper motion star is as follows. (1) The
code identified the plate feature in the POSS-I or POSS-II image that
is responsible for the residual feature. (2) The plate feature is
removed from the POSS image in which it is found. (3) The modified
plate image, lacking the plate feature, is then subtracted from the
second plate image. (4) The plate feature is then matched to any
plate feature remaining on the second plate image after the
subtraction. Because the first plate image is complete except for the
now removed candidate moving feature, subtracting it from the second
plate image removes everything except the counterpart of
the moving object on the second plate image. (5) A probability of
match is calculated based on the similarity between the two
features. The code also calculates the proper motion and estimates the
magnitude of the candidate high proper motion star.

The detailed algorithm has been tested and specifically adapted to
work best on scans of photographic plates, and it achieves extremely
high detection rates. A significant problem arises when a HPM star
moves into the image of a brighter star on one plate (when this
happens, we say that the high proper motion star passes ``in front
of'' the brighter star). When this occurs, the residual image of the
high proper motion star is assigned by the code to the brighter star
because the code has no way of knowing that the residual image is due
to a proper motion star passing in front of the bright
star. Fortunately, HPM events always leave {\em two} distinct traces
in the residual images, so if one of the residual images corresponds
to the image of the HPM object passing in front of a brighter star,
the other one will (most of the time) be correctly identified. The
code is usually able to make the match with the counterpart
passing in front of the brighter star. This is actually one of the
greatest strengths of the code: {\em the ability to correctly identify a
high proper motion star even if it passes in front of a brighter star
on one of the plates}. Several of the new high proper motion stars
that we have identified with SUPERBLINK occur in such a fashion.

The only case when the code will definitely miss an HPM star is when
the star is passing in front of brighter stars in {\em both} the
POSS-I and POSS-II plates. Admittedly, this failure of the code could
hardly be noticed since such a star is also very likely to escape
detection even by careful visual inspection of the plates. However,
this is an important issue because we do expect that a small number of
faint, high proper motion stars will escape detection, especially in
low galactic latitude fields where crowding can be significant.

The two parameters that determine the computing time used by the
SUPERBLINK code are the lower intensity limit for the detection of
faint residual features and the search radius for a matching
counterpart. Experiments have shown that for the combination of POSS-I
and POSS-II plates, optimal results are obtained by limiting the
identification of variable features with magnitude brighter than about
r=20, and using a search radius smaller than $2\arcmin$. Using these
parameter values, a one square degree area can be processed in about
15 minutes on a Linux workstation with dual Pentium-III
microprocessors. Over 10,000 CPU-hours have been invested so far in
searching the northern sky for new HPM stars.

\subsection{Estimation of POSS-II magnitudes}

The SUPERBLINK software provides POSS-II red magnitude estimates
for the detected high proper motion stars by simple calculation of the
plate flux. Because we only retrieve the red plates, we do not have any
direct means to get colors for the objects. But since the second
version of the Guide Star Catalog (GSC-II) catalog has been made
publicly available, we have been searching for counterparts of
our high proper motion stars in the public version of the GSC-II (the
GSC2.2.1 catalog) to obtain POSS-II $r$ and $b$ colors. 

The GSC2.2.1 is based on scans of the photographic Sky Survey plates,
exactly the same material that we are using for our survey, and can be
regarded as an improved version of the Guide Star Catalog (Lasker {\it et
al.} 1990, Russell {\it et al.} 1990, and Jenkner {\it et al.}
1990). The GSC-II is maintained by the Space Telescope Science
Institute, and the GSC2.2.1 is available electronically through the
ViZier service of the Centre de Donn\'ees Astronomiques de Strasbourg
(catalog number I/271).

All entries in the GSC2.2.1 are provided with coordinates and epochs,
which correspond to the epoch of the photographic plate when the star
was observed. For each GSC2.2.1 star located within 1$\arcmin$ of one
of our HPM stars, we calculate the astrometric distance between the
location of the GSC2.2.1 star and the estimated location of the HPM
star candidate {\em at the epoch of the GSC2.2.1 star} (recall that
our catalog quotes the position of the HPM stars at the 2000.0
epoch).

While cross-correlating our sample of high proper motion stars into
the GSC2.2.1 catalog, we found that $\sim25\%$ of the stars could be
matched with two or more different objects in the GSC2.2.1
catalog. Every time this happened, the different objects listed in the
GSC2.2.1 had different quoted epochs of detections and thus obviously
corresponded {\em to the same star observed in different POSS-II
plates at different epochs}. In most cases, one object has a quoted
red magnitude but no blue magnitude, while the other has a blue
magnitude, but no red magnitude. This is because the same star has
been observed on plates of different colors at different
times. Whenever a star has been observed twice in the same color and
quoted as two different objects, we have used the mean of the quoted
magnitudes.

A small number of stars were found not to have any GSC2.2.1
counterpart. The majority of these are stars which are beyond the
current magnitude limit of the Second Guide Star Catalog (r=18.5), but
a handful of brighter stars could still not be found. The majority of
those missing stars are moderately bright stars that just happened to
be passing close to brighter stars at the epoch when the POSS-II image
was obtained. They can be easily identified from the POSS-I plates,
but their identification is difficult on the POSS-II plates, on which
the GSC2.2.1 is based. In one specific case, the missing star was
found on the POSS-II image blended with a bright track left by the
passage of an airplane; it could not possibly have been identified as
a star from the POSS-II scans alone.

We use in our catalog the quoted GSC2.2.1 $r$ and $b$ magnitudes for
all the stars we were able to match from GSC2.2.1. For all other
stars, we use the $r$ magnitude estimated by SUPERBLINK. The GSC2.2.1
magnitudes are a significant improvement over the magnitudes quoted in
the Luyten proper motion catalogues (see \S4.3 below).


\section{Inclusion of bright high proper motion stars from the
Tycho catalogs}

The SUPERBLINK software does not work efficiently on stars which are
extremely saturated on the DSS images. SUPERBLINK works best for stars
fainter than magnitude $r=12$ and cannot handle stars brighter than
about $r=9$ at all. Fortunately, the range of magnitudes over which the
code is much less efficient corresponds to the faint limit of the stars
which have been measured during the HIPPARCOS astrometric
mission, and listed in the Hipparcos and Tycho catalog (ESA 1997) and
in the Tycho-2  catalog (Hog {\it et al.} 2000). Provided that the
SUPERBLINK and Tycho samples overlap over a significant range in
magnitudes, the two surveys are really complementary, and we thus have
a means to build a complete census of all high proper motion stars
down to r=20.

Our recovery rate of high proper motion Tycho stars in our survey
area is plotted in Figure 2 as a function of the HIPPARCOS V
magnitude. Not a single star in the V=8 magnitude bin was identified
by SUPERBLINK, which sets the absolute upper limit of our code in
handling bright stars. The efficiency rises rapidly as the stars get
fainter, and the recovery rate reaches over 80\% in the V=10 magnitude
bin. The SUPERBLINK software was fully successful in recovering the
stars in the last magnitude bin of the Tycho catalogs (V=12). However,
this last bin contains only 15 stars, and while this does suggest a
high recovery rate, we cannot affirm that the code is actually more
than about 90\% efficient in that range. 

Nevertheless, this does suggest a good complementarity between the
Tycho catalogs and our survey. The combination of high proper motion
stars from the Tycho catalogs and from our survey should provide a
virtually complete census of {\em all} high proper motion stars in the
survey area down to the magnitude limit of our survey. The main
limitation lies in that the HIPPARCOS satellite relied on an input
catalog, and only observed a subsample of selected stars fainter than
V$\approx9$. Because our DSS-based survey is incomplete for stars in
the $9<$V$<12$ range, there is the possibility that some high proper
motion stars in that range are not in the Tycho catalogs and will not
have been recovered in our survey. The acid test to check if the
combined Tycho/DSS catalog is complete throughout the full range of
magnitudes is a comparison with the previous most complete catalog of
high proper motion stars: the Luyten catalogs.


\section{Comparison with the Luyten proper motion catalogs}

\subsection{Astrometric errors in the Luyten catalogs: rebuilding the
Luyten proper motion survey}

We find that there are numerous, significant errors in the stellar
positions quoted in the LHS and in the NLTT catalogs. The LHS and NLTT
catalogs are supposed to be essentially redundant for stars with
proper motions $\mu>0.5\arcsec$ yr$^{-1}$. The main difference between
the two is that the NLTT catalog also lists stars down to
$\mu>0.18\arcsec$ yr$^{-1}$. In all appearances, the differences are
only minor: the LHS catalogs lists 3561 stars with $\mu>0.5\arcsec$
yr$^{-1}$ while the NLTT lists 3555. While both the LHS and NLTT give
R1950 coordinates for all entries, the LHS catalog also gives J2000
coordinates.

The R1950 coordinate should be understood as the coordinate of the
star at epoch 1950.0 and equinox 1950. The J2000 coordinate refers to
the position of the star at epoch 2000.0 in the equatorial coordinate
system with equinox 2000. We should expect to be able to re-derive the
quoted J2000 position by adding the quoted proper motion of the
star to the R1950 position and precessing the result. Using the
electronic version of the catalog obtained from VIZIER, we have
attempted to reproduce this result. We were quite surprised to find
that a large number of stars had significant differences between the
J2000 coordinate that we calculated based on the quoted R1950
coordinate and proper motion, and the J2000 coordinate quoted by
Luyten. Of a total of 4421 stars, 570 showed discrepancies of over
20$\arcsec$, 147 showed discrepancies of over 1$\arcmin$,
and 84 showed discrepancies of over 5$\arcmin$. We find that in
the majority of the cases, it is the quoted J2000 coordinate that is
clearly in error. We verified this by matching the stars with large
errors with the updated LHS catalog compiled by Bakos, Sahu, and
N\'emeth (2002). All the stars for which we found a very large
discrepancy between the J2000 and R1950 position are those that Bakos,
Sahu, and N\'emeth were unable to recover in their investigation. We
can therefore identify the main source of their many problems at
recovering LHS stars in the DSS with the fact that {\em they used the
error-filled J2000 column as their source of coordinates in the LHS
catalog}.

It is most unfortunate that the searchable electronic version of
the LHS catalog found at CDS uses the J2000 column as the main source
of coordinates. Even worse, the SIMBAD database also uses the
erroneous column as the main source of coordinates for most objects
that are LHS stars. This leads to a number of HPM stars being quoted
twice in the Simbad database, with one entry for the LHS star at the
erroneous location and another, separate entry for the same star,
quoted under its other known names at the accurate location. We
conclude that coordinates quoted in the LHS catalog are extremely
unreliable, and that they should be excluded from use in the SIMBAD
astronomical database. In contrast, we found the coordinates retrieved
from the electronic version of the NLTT catalog to be significantly
more reliable.

We have thus compiled a more accurate list of Luyten stars by
merging the information provided by the LHS and NLTT. Unfortunately,
the LHS and NLTT have different numbering systems, and the only means
to match an NLTT star with an LHS star is to match their coordinates.
We thus cross-correlated all entries in the LHS and NLTT, using both
the R1950 and the (precessed) J2000 coordinates from the LHS, and the
R1950 coordinates for the NLTT. Given the sometimes large difference
between the positions, we matched the stars by performing several
passes using increasingly larger search radii (up to $5\arcmin$)
until most of the LHS and NLTT stars could be paired. We found a total
of 3554 stars with $\mu>0.5\arcsec$ yr$^{-1}$ listed in both the LHS and
NLTT catalog. In the process, we discovered 6 stars which were listed
twice in the LHS catalog under different names (LHS28 and LHS201,
LHS1611 and LHS5086, LHS2785 and LHS2786, LHS5126a and LHS6132,
LHS5215a and LHS6231, LHS5217a and LHS6235). While the LHS R1950
coordinates were generally found to provide the best match to the NLTT
counterparts, 411 LHS stars could be matched better with NLTT objects
when we used their precessed J2000 coordinates. We generally found
good agreement between the quoted R1950 positions in the LHS and
NLTT. However, most of the NLTT coordinates appear to have been copied
directly from the LHS, so this is no guarantee that the locations are
accurate.

There were 32 LHS stars with $\mu>0.5\arcsec$ yr$^{-1}$ for which we
could not find an NLTT counterpart with the automated procedure, and
53 NLTT stars with $\mu>0.5\arcsec$ yr$^{-1}$ which we could not match
with an LHS star. Close examination of all these entries revealed that
28 more stars could be matched together by assuming that one of the
quoted coordinates (either from the LHS or from the NLTT) had an error
larger than $5\arcmin$. In all these cases, the LHS and NLTT entries
have the same quoted proper motion, proper motion angle, and
magnitude, but different quoted positions. We searched for those high
proper motion stars in the DSS at both quoted positions. We were able
to recover all the high proper motion stars at one of the quoted
positions, therefore identifying which ones were in error (see Table
1). For 21 of the stars, it appears to be a simple transcription error
affecting one digit or a sign, always in the declination column except
for LHS3415/NLTT47039 where the error in in the RA column. For the
other 7 stars, there is a discrepancy $\approx15\arcmin$ in the
declination which affects only NLTT entries.

In the end, we were left with 4 LHS stars with $\mu>0.5\arcsec$
yr$^{-1}$ for which we could not find any counterpart in the NLTT and
25 NLTT stars with $\mu>0.5\arcsec$ yr$^{-1}$ for which we could not
find any counterpart in the LHS. We verified the existence of all
these objects by looking in DSS fields at the quoted locations. We
confirmed the existence of all 4 LHS stars which are missing from the
NLTT (see Table 2). We also confirmed the existence of 22 on the NLTT
stars absent from the LHS, but were unable to recover NLTT11999,
NLTT22764, and NLTT46584 within $5\arcmin$ of their quoted
location. Furthermore, we found that 4 of the recovered NLTT stars had
quoted proper motions overestimated by a factor 10, and are actually
stars with $\mu<0.5\arcsec$ yr$^{-1}$ (see Table 3). This means that
18 NLTT stars with $\mu<0.5\arcsec$ yr$^{-1}$ were actually omitted
from the LHS catalog. Overall, {\em the complete sample of
$\mu<0.5\arcsec$ yr$^{-1}$ stars compiled by Luyten includes 3576
distinct objects}. We used this verified compilation of high proper
motion Luyten stars to match the stars found by SUPERBLINK or listed
in the Tycho catalogs. Within our survey area ($\delta>-2.7^{\circ},
-25^{\circ}<|b|<+25^{\circ}$) there is a total of 674 stars from the
Luyten sample.

\subsection{Recovery of LHS and NLTT stars}

We have cross-correlated all entries in the combined list of HPM stars
that we generated from SUPERBLINK and the Tycho catalogs with the
updated Luyten master catalog reconstructed from the LHS and NLTT. A
first pass was made where we matched stars with positions similar to
within $1\arcmin$. We made successive passes with gradual increases in
the search radius, always requiring that matching objects have proper
motions similar to within 10\% and proper motion vectors similar to
within $10^{\circ}$. We found that 85\% of the Luyten stars in the
area covered by our survey could be matched within 10$\arcsec$ of
stars found by SUPERBLINK (see Figure 3). By increasing the search
radius to 1$\arcmin$, we could match 665 of the Luyten stars (all but
9 of them). One more star was matched (LHS6374) which we found at a
distance of $563\arcsec$ of the location quoted by Luyten (the
matching star has a very similar magnitude and proper motion). We
could not find any counterpart in our catalog for the remaining 9
stars (see \S5 below).

Figure 4 presents the complete results of our analysis. We use a color
scheme to identify four classes of objects. Black: HPM stars that were
found in the Tycho catalogs. Blue: HPM stars found by SUPERBLINK in
the DSS, and matched to previously known HPM stars, listed in the
Luyten catalogs. Red: HPM stars listed in the Luyten catalog which
are not listed in the Tycho catalogs and could not be recovered with
SUPERBLINK. Green: HPM stars identified in the DSS with SUPERBLINK but
that do not correspond to any entry in the Luyten catalogs, i.e. newly
discovered HPM stars. The histogram clearly illustrates the fact
that the HIPPARCOS satellite has only looked at the ``tip of the
iceberg'' of HPM stars with $\mu>0.5 \arcsec$ yr$^{-1}$, since the bulk
of the stars are in the $10<r<16$ magnitude range. As expected, most
of the new discoveries are at the faint end of the distribution. The
10 Luyten stars that have been ``missed'' actually include 3 stars
which have proper motions too large to be detected by SUPERBLINK
($\mu>2.0\arcsec$ yr$^{-1}$). Three more ``missed'' Luyten stars are
close to the bright star limit for SUPERBLINK, while also being below
the completeness limit of the Tycho catalog.

Based on these results, we have calculated the completeness of the
Tycho and Luyten catalogs, and estimated the completeness of our own
survey as a function of magnitude; the results are displayed in Figure
5. To calculate the completeness of the Tycho and Luyten catalogs, we
have simply divided, for each magnitude bin, the number of HPM stars
found in those catalogs by the total number of HPM stars known (those
now listed in the present catalog, which include the new stars found
by SUPERBLINK). To estimate the completeness of our own survey, we
calculated the recovery rate of Luyten stars for each magnitude bin.
We find that, at low galactic latitudes, the Luyten catalog is
complete only down to $r$=13, and the completeness falls below
50\% for stars fainter than $r$=15. The Tycho catalog is complete for
HPM stars only down to $r=8$ and is very incomplete below $r=12$. The
Tycho-2 catalog is thus a good complement to SUPERBLINK, which is
extremely efficient, but only for stars fainter than $r$=12. Note that
the $r=17$ and $r=18$ bins give a 100\% recovery rate but are based on
the recovery of only 15, and 6 Luyten stars, respectively (all of them
recovered by SUPERBLINK). While this is consistent with a very high
efficiency for SUPERBLINK, we cannot exclude the possibility that the
faint Luyten stars, having been discovered with more limited methods,
are simply the most obvious HPM stars to be found, and that more
elusive objects have actually been missed by SUPERBLINK. However,
based on our recovery of NLTT stars with proper motions in the range
$0.18<\mu<0.50\arcsec$ yr$^{-1}$ (analysis in progress), we find that
SUPERBLINK is indeed highly efficient in finding HPM stars down to
$r=19$.

Although there appears to be a concentration of new stars at very low
galactic latitudes ($|b|<10^{\circ}$) some new stars have been
discovered even at $|b|>20^{\circ}$, which indicates that our analysis
of higher galactic latitudes (under way) is likely to discover still
more new faint HPM stars.

\subsection{Magnitude and color differences}

We compare the red and blue magnitudes and color as quoted
in the NLTT and LHS catalogs (the Luyten-r and Luyten-b magnitudes),
and the magnitudes and colors estimated from our survey with the help
of the GSC2.2.1 ($r$, $b$, and $b-r$), to the magnitudes and colors of
the same stars that are found in the Tycho catalogs ($B_T$, $V_T$,
$B_T-V_T$). The Tycho catalog magnitudes are accurate to about 0.1
mag, while photographic magnitudes are generally accurate to only 0.5
mags. We thus use the Tycho magnitudes as a standard of reference to
compare the accuracy of the Luyten and POSS-II magnitudes.

Figure 6 reveals that the POSS-II magnitudes are significantly more
accurate than the magnitudes estimated by Luyten. It also shows
that Luyten-r, the ``red'' magnitude quoted in the Luyten
catalogs, is actually closer to a visual V magnitude; Luyten-b and
Luyten-r do match $B_T$ and $V_T$ relatively well. This warrants
the use of the new POSS-II magnitudes as the reference for our updated
catalog of HPM stars.

Figure 7 shows the generally good agreement between the Luyten-r -
Luyten-b color and the Tycho $B_T-V_T$ color, while there is
a substantial difference between these two and the POSS-II $b-r$
color. The Tycho $B_T$ and $V_T$ passbands are spaced further apart
than Johnson $B$ and $V$; there is a relationship
$B-V=0.85*(B_T-V_T)$. Figure 7 also shows that there is an extremely
large scatter in the color-color relationship between the Luyten and
POSS-II magnitude. Since we have shown that the POSS-II magnitudes are
more accurate than the Luyten estimated magnitudes, we suggest that
any classification based on the old Luyten magnitudes should be
revised. It is possible that {\em stars suspected to be white dwarfs
based on the Luyten photometry may turn out to be red dwarfs and
vice-versa}. We demonstrate this by reconstructing the reduced proper
motion diagram for the Luyten stars in the area covered by our survey.

\subsection{Reduced proper motions}

The reduced proper motion $H$ is obtained by adding the logarithm
of the proper motion to the apparent magnitude of the star (Luyten
1925). Because both the apparent magnitude and proper motions are a
function of the distance to the star, the distance terms
cancel, and one is left with a measure of the combined absolute
magnitude and transverse velocity. The reduced proper motion can be
defined in a specific passband e.g.:
\begin{displaymath}
H_{r} = r + 5 + 5 \log{\mu} = M_r + 5 \log{v_t} -3.38 ,
\end{displaymath}
where $r$ and $\mu$ are the measured apparent magnitude and proper
motion (in $\arcsec$ yr$^{-1}$) and $M_r$ and $v_t$ are the absolute
magnitude and sky-projected velocity component (in km s$^{-1}$). The
reduced proper motion is useful when plotted against color, as it
yields a diagram similar to the Hertzprung-Russel diagram, but with an
additional scatter due to the $\log{v_t}$ term. Stars of a given color
with unusually large reduced proper motion either are intrinsically
sub-luminous (such as white dwarfs) and/or have unusually large
transverse velocities (halo subdwarfs). The main use of the reduced
proper motion diagram for faint stars is thus to separate out the
white dwarfs and the high velocity stars from the local K and M
dwarfs.

We have built reduced proper motion / color diagrams for all the stars
in our catalog for which we had Luyten-r and Luyten-b magnitudes and
GSC2.2.1 $r$ and $b$ magnitudes. These essentially represent the
subsample of old Luyten stars which were faint enough that we could
get POSS-II magnitudes (brighter stars are saturated on the POSS
plates and their magnitudes are uncertain). This excludes from the
sample most of the LHS and NLTT stars observed by HIPPARCOS. The
purpose of this exercise is to compare the quality of the red and blue
magnitudes quoted by Luyten and those derived from the GSC2.2.1. A
comparison of the reduced proper motion diagrams attests to the
generally higher quality of the POSS-II magnitudes (see Figure 8). The
scatter in color is smaller, and the white dwarf sequence (to the
lower left) is better separated from the K-M dwarf sequence (the bulk
of the stars on the upper right). The better separation is achieved
partly because of the higher quality of the color determination in the
POSS-II data, but also because the $b-r$ color
index spans a larger dynamic range than Luyten-b - Luyten-r (since the
Luyten color is actually equivalent to a $B-V$). The larger span in
dynamic range insures that the red dwarfs are shifted more to the left
of the diagram, where they are more easily separated from the bluer
white dwarfs.


\section{Luyten stars not observed by HIPPARCOS and missed by
SUPERBLINK}

A total of 9 stars listed in the Luyten catalogs could not be found
either in the Tycho catalog or on the DSS plates with
SUPERBLINK. Three of the Luyten stars for which we could not find a
match (LHS18, LHS35, LHS64) exceed the 2.0$\arcsec$ yr$^{-1}$ search
limit of SUPERBLINK, and are also not entered in the Tycho-2
catalog. We manually recovered these stars by inspecting the DSS
fields in which they would be placed according to Luyten's
coordinates, and they were recovered at Luyten's predicted positions. 

We also examined DSS fields centered on the quoted location for the 6
other Luyten stars which SUPERBLINK failed to recover. We were able to
visually identify 5 of the stars. Three of these (LHS223, LHS464,
LHS3489) are relatively bright $r$=12-14 magnitude stars which are at
the bright star limit of SUPERBLINK and below the completeness limit
of the Tycho catalog. The other two stars (LHS153, LHS1103) are
$r=16-17$ stars, normally within the detection range of SUPERBLINK,
but were missed for reasons that we have not fully explained yet. Both
objects lie close to brighter stars at one epoch, but SUPERBLINK
normally handles similar cases very well. One possibility we are
currently investigating is that difficulty in handling the very
different PSF in the POSS-I and POSS-II images for those specific
objects might be the source of the problem.

The only LHS star which we were absolutely unable to recover,
either with our software or by visual examination of the DSS plates,
is the very faint star LHS1657, quoted by Luyten as having a red
magnitude r=19.0 and a blue magnitude b=21.0. Absolutely no star could
be found at the location predicted by Luyten, and visual examination
of a $15\arcmin\times15\arcmin$ field centered of the quoted position
for this star failed to recover any HPM star at all. We conclude
that, in all likeliness, this is a bogus entry in the LHS catalog (the
only one in our survey area).

Overall, we find that SUPERBLINK achieves a recovery rate exceeding
99.5\% for faint Luyten stars. This suggests that not more that a
handful of HPM stars have been missed by our software, at least down
to $r$=19.


\section{New high proper motion stars discovered by SUPERBLINK}

We have identified a total of 147 stars which did not match any entry
from the Luyten catalogs. Six of those are close companions to Luyten
stars (listed as single in the LHS and NLTT) identified with
HIPPARCOS, and listed in the Tycho-2 catalog. The other 141 stars are
entirely new objects discovered with SUPERBLINK, and identified here
for the first time. 

We verified if any of the new stars found by SUPEBLINK could be a
previously unreported common proper motion companion to a Luyten star.
We checked for the presence of Luyten stars within $30\arcmin$ of the
new stars, with proper motion magnitudes within 5\%. We discovered 2
new faint stars which are common proper motion companions to brighter
Luyten stars (see \S7 below).

The names, coordinates, and magnitudes of the 141 stars are listed in
Table 4. We give the estimated coordinates of the stars for the epoch
2000.0, which we extrapolated from the measured position on the
POSS-II with the estimated proper motion vector. The magnitude $pm$ of
the proper motion, and the direction of the proper motion vector
(proper motion angle, or $pma$) are also given in the table. We give
the estimated values of the red and blue magnitudes ($b$ and $r$) in
the POSS-II system. We are quoting the r and b magnitudes from the
GSC2.2.1 for all those stars for which we were able to obtain a
match. For all other stars (mainly stars with $r<18.5$) we estimated
the magnitudes ourselves from the integrated plate flux. We could not
obtain an estimate of the b magnitude for several of the faintest
stars; these were too faint on the blue plates and often did not have
any observable blue counterpart. All the quoted magnitudes are
accurate to $\pm0.5$. Finding charts for all 141 stars are presented
in the appendix.

The distribution of new and old stars with galactic latitude is shown
in Figure 9, for stars in the range $13<r<16$ (moderately faint) and
$16<r<20$ (faint). The Luyten survey was virtually complete for bright
stars ($r<13$). A significant number of moderately faint stars were
missing at low galactic latitudes; most of the new moderately faint
HPM stars were found within 15$^{\circ}$ of the plane. On the other
hand, Luyten's census of faint HPM stars was significantly incomplete,
even at relatively high galactic latitudes ($|b|\approx20^{\circ}$), and
listed only a handful of objects within $10^{\circ}$ of the galactic
plane. It remains to be verified whether significant numbers of faint
HPM stars can be discovered at higher galactic latitudes
($|b|>25^{\circ}$). The distribution with galactic latitude of the
updated census of HPM stars is now consistent with a uniform
distribution, which suggests that the survey is now complete, at least
down to R=19.


\section{New common proper motion companions to LHS stars}

\subsection{LSR0013+5437: a companion to LHS1039}

We have discovered an r=17 star LSR0013+5437 located 343$\arcsec$ to
the southeast of the r=13 star LSR0012+5439 (=LHS1039). Both stars
have a proper motion $\mu\approx0.98\arcsec$ yr$^{-1}$, with a proper
motion angle $\approx52^{\circ}$. The primary star, also known as
G217-38 (see Giclas, Burnham, \& Thomas 1971), is a suspected M dwarf
with no published spectrum or spectroscopically determined spectral
type. A parallax of $\sim0.027\arcsec$ for LHS1039 has been measured
by Harrington \& Dahn (1980), which suggest the star is at a distance
$\sim37$pc. At that distance, the projected separation between the
components is $\approx12,500$AU.

\subsection{LSR0401+5131: a companion to LHS1618}

We have discovered an r=16 star LSR0401+5131 located 495$\arcsec$ to
the north of the r=12 star LSR0401+5123 (=LHS1618). Both stars have a
proper motion $\mu\approx0.85\arcsec$ yr$^{-1}$, with a proper motion
angle $\approx-24^{\circ}$. The star LHS1618 has been
spectroscopically observed and classified M3.5 by Reid, Hawley, \&
Gizis (1995). It is a suspected nearby star listed in the {\it Catalog
of Nearby Stars, Preliminary 3rd Version} (Gliese \& Jahreiss 1991),
with an estimated distance of $\approx39$pc. At that distance, the
projected separation between the components is a large
$\approx19,300$AU. The 4 magnitudes difference in their luminosity
suggests that the secondary is an M5-M6 dwarf.

\subsection{LSR1833+2219: a companion to LHS3392}

This new distant companion to the nearby flare star LHS3392 (=V774
Her, =GJ718) has already been presented and discussed in a previous
paper by us (L\'epine, Shara \& Rich 2002). Please refer to that paper
for more details.


\section{Conclusions}

The current census of northern hemisphere stars with proper motion
$0.5<\mu<2.0 \arcsec$ yr$^{-1}$ and within 25$^{\circ}$ of the
galactic plane is now essentially complete down to $r=19$. There are
at most a handful of such objects which remain to be identified.

The Luyten catalogs of high proper motion stars (LHS, NLTT) are
significantly incomplete for faint ($R>15$) stars at low galactic
latitudes. The Luyten catalogs also contain numerous errors in the
positions of the stars, and should be replaced with new a new catalog
of high proper motion stars.

We are currently expanding our search for high proper motion stars at
higher galactic latitude in the northern sky. Results will be
published in the next paper of this series.


\acknowledgments

{\bf Acknowledgments}

The late Barry Lasker was, for many years, project scientist of the
Guide Stars Catalogs, and the Digitized Sky Surveys, at the Space
Telescope Science Institute. The enormously positive impacts of these
projects will continue to be felt in astronomy for decades to
come. This work (and others to follow) are only possible because of
Barry's vision, tenacity and dedication, and we are grateful for his
friendship and astronomical legacy. We also gratefully acknowledge the
enormous efforts of Brian McLean and dozens of his colleagues:
astronomers, engineers, programmers, data and archive technicians and
plate scanners who made the Digitized Sky Surveys a reality.

This research program has been supported by NSF grant AST-0087313 at
the American Museum of Natural History, as part of the NSTARS program.

This work has been made possible through the use of the Digitized Sky
Surveys. The Digitized Sky Surveys were produced at the Space
Telescope Science Institute under U.S. Government grant NAG
W-2166. The images of these surveys are based on photographic data
obtained using the Oschin Schmidt Telescope on Palomar Mountain and
the UK Schmidt Telescope. The plates were processed into the present
compressed digital form with the permission of these institutions. The
National Geographic Society - Palomar Observatory Sky Atlas (POSS-I)
was made by the California Institute of Technology with grants from
the National Geographic Society. The Second Palomar Observatory Sky
Survey (POSS-II) was made by the California Institute of Technology
with funds from the National Science Foundation, the National
Geographic Society, the Sloan Foundation, the Samuel Oschin
Foundation, and the Eastman Kodak Corporation. The Oschin Schmidt
Telescope is operated by the California Institute of Technology and
Palomar Observatory. The UK Schmidt Telescope was operated by the
Royal Observatory Edinburgh, with funding from the UK Science and
Engineering Research Council (later the UK Particle Physics and
Astronomy Research Council), until 1988 June, and thereafter by the
Anglo-Australian Observatory. The blue plates of the southern Sky
Atlas and its Equatorial Extension (together known as the SERC-J), as
well as the Equatorial Red (ER), and the Second Epoch [red] Survey
(SES) were all taken with the UK Schmidt. 

This work has been made possible through the use of the Guide Star
Catalogue-II. The Guide Star Catalogue-II is a joint project of the
Space Telescope Science Institute and the Osservatorio Astronomico di
Torino. Space Telescope Science Institute is operated by the
Association of Universities for Research in Astronomy, for the
National Aeronautics and Space Administration under contract
NAS5-26555. The participation of the Osservatorio Astronomico di
Torino is supported by the Italian Council for Research in
Astronomy. Additional support is provided by European Southern
Observatory, Space Telescope European Coordinating Facility, the
International GEMINI project and the European Space Agency
Astrophysics Division.

The data mining required for this work has been made possible with the
use of the SIMBAD astronomical database and VIZIER astronomical
catalogs service, both maintained and operated by the Centre de
Donn\'ees Astronomiques de Strasbourg (http://cdsweb.u-strasbg.fr/).

\appendix

\section{Finder charts}

Finder charts of high proper motion stars are been generated as a
by-product of the SUPERBLINK software. We present here finder charts
for all the new, low galactic latitude, high proper motion stars
presented in this paper. All the charts consist in pairs of
$4.25\arcmin\times4.25\arcmin$ images showing the local field at two
different epochs. The name of the star is indicated in the center just
below the chart, and correspond to the name given in Table 1. To the
left is the POSS-I field, with the epoch of the field noted in the
lower left corner (typically in the 1953-1955 range). To the right is
the modified POSS-II field which has been shifted, rotated, and
degraded in such a way that it matches the quality and aspect of the
POSS-I image. The epoch of the POSS-II field is noted on the lower
right corner. High proper motion stars are identified with circles
centered on them.

The charts are oriented in the local X-Y coordinate system of the
POSS-I image; the POSS-II image has been remapped on the POSS-I
grid. This means that north is generally up and east left, but the
fields might appear rotated by a small angle for high declination
objects. Sometimes a part of the field is missing: this is an artifact
of the code. SUPERBLINK works on $17\arcmin\times17\arcmin$ DSS
subfields. If a high proper motion star is identified near the edge of
that subfield, the output chart will be truncated. Similar finder
charts are automatically generated for every single object identified
by the code, and we are currently building a large electronic catalog
of two-epoch finder charts.


\begin{deluxetable}{rrrrrrrrrrrr} 
\tablecolumns{12} 
\tablewidth{0pc} 
\tabletypesize{\scriptsize}
\tablecaption{Retranscription errors in the Luyten catalogs
\tablenotemark{a}}
\tablehead{ 
\multicolumn{6}{c}{LHS} & \multicolumn{6}{c}{NLTT} \\
\cline{2-6} \cline{8-12} \\
\colhead{LHS\tablenotemark{b}} & 
\colhead{RA1950} &
\colhead{DE1950} &
\colhead{r} & 
\colhead{pm($\arcsec$ yr$^{-1}$)} & 
\colhead{pma($^{\circ}$)} &
\colhead{NLTT\tablenotemark{c}} & 
\colhead{RA1950} &
\colhead{DE1950} &
\colhead{r} & 
\colhead{pm($\arcsec$ yr$^{-1}$)} & 
\colhead{pma($^{\circ}$)}
}

\startdata 
1092 & 00 29 18 &(-31 04.2)& 13.6 & 0.521 & 182.1 & 1706 & 00 29 18 & -30 04.2 & 13.6 & 0.521 & 182 \\
1266 & 01 32 59 & +41 22.9 & 13.2 & 0.570 &  92.7 & 5312 & 01 32 59 &(+41 38.2)& 13.2 & 0.570 &  92 \\
1382 & 02 15 25 & +01 31.3 &  6.0 & 0.532 &  44.5 & 7585 & 02 15 25 &(+01 45.4)&  6.0 & 0.532 &  45 \\
 160 & 02 51 10 & -63 54.5 & 11.0 & 1.149 &  58.4 & 9248 & 02 51 12 &(-63 42.0)& 11.0 & 1.150 &  58 \\
1770 & 05 36 46 & -46 07.6 &  7.5 & 0.508 & 197.0 &15426 & 05 36 48 &(-47 08.0)&  7.6 & 0.510 & 197 \\
2053 & 08 46 02 & +36 42.9 & 10.9 & 0.521 & 205.9 &20295 & 08 46 02 &(+36 47.9)& 10.9 & 0.521 & 206 \\
 255 & 08 52 33 & +01 45.1 &  9.6 & 1.066 & 176.7 &20528 & 08 52 33 &(+01 32.7)&  9.6 & 1.066 & 177 \\
2301 & 10 40 36 &(+23 48.0)& 14.4 & 0.517 & 185.1 &25146 & 10 40 36 & +29 48.0 & 14.4 & 0.517 & 185 \\
2384 & 11 14 04 &(+44 44.9)& 14.0 & 0.520 & 268.1 &26831 & 11 14 04 & -43 44.9 & 14.0 & 0.520 & 268 \\
2401 & 11 21 29 &(+18 05.3)& 12.9 & 0.616 & 266.7 &27268 & 11 21 29 & -18 05.3 & 12.9 & 0.616 & 267 \\
2417 & 11 28 19 & -56 51.5 &  8.9 & 0.565 & 272.1 &27605 & 11 28 18 &(-57 52.0)&  8.9 & 0.560 & 272 \\
2526 & 12 10 40 &(+12 06.5)&  7.8 & 0.606 & 177.0 &30006 & 12 10 40 & +11 06.5 &  7.8 & 0.606 & 177 \\
2532 & 12 12 18 & +62 33.0 & 18.3 & 0.508 & 266.7 &30091 & 12 12 18 &(+62 18.3)& 18.3 & 0.508 & 267 \\
2848 & 13 59 41 & -20 45.9 & 12.8 & 0.628 & 125.3 &36049 & 13 59 41 &(-26 45.9)& 12.8 & 0.628 & 125 \\
2946 & 14 33 29 & +67 21.4 & 13.8 & 0.611 & 297.9 &37844 & 14 33 29 &(+67 08.6)& 13.8 & 0.611 & 298 \\
3021 & 15 02 07 &(-47 43.4)& 16.2 & 0.547 & 185.4 &39208 & 15 02 07 & -17 43.4 & 16.2 & 0.547 & 185 \\
3047 & 15 13 55 & -29 35.8 & 15.2 & 0.932 & 248.5 &39742 & 15 13 55 &(-28 21.0)& 15.2 & 0.932 & 249 \\
3286 & 17 20 35 &(+32 11.8)& 11.9 & 0.596 & 198.0 &44684 & 17 20 35 & -32 11.8 & 11.9 & 0.596 & 198 \\
3415 & 18 47 33 & +11 15.5 & 13.8 & 0.889 & 208.0 &47039 &(18 46 33)& +11 15.5 & 13.8 & 0.889 & 208 \\
3492 & 19 47 49 & -51 01.8 & 15.9 & 0.852 & 191.3 &48302 & 19 47 48 &(-50 02.0)& 15.9 & 0.850 & 191 \\
3498 & 19 52 44 & -27 25.6 & 15.0 & 0.608 & 184.5 &48425 & 19 52 44 &(-22 25.5)& 15.6 & 0.608 & 185 \\
3587 & 20 46 53 & +15 54.6 & 13.1 & 0.559 & 207.5 &49948 & 20 46 53 &(+16 54.6)& 13.1 & 0.559 & 208 \\
  66 & 21 14 20 & -39 03.7 &  6.4 & 3.453 & 250.5 &50917 & 21 14 20 &(-38 03.7)&  6.4 & 3.453 & 251 \\
3674 & 21 22 20 & -65 35.6 &  4.3 & 0.804 &   5.8 &51220 & 21 22 18 &(-54 36.0)&  4.3 & 0.800 &   6 \\
3743 & 21 58 53 & +26 55.1 & 14.0 & 0.598 &  93.9 &52695 & 21 58 53 &(+28 55.1)& 14.0 & 0.598 &  94 \\
3809 & 22 23 43 & +02 45.1 & 15.8 & 0.663 & 228.4 &53835 & 22 23 43 &(+03 45.1)& 15.8 & 0.663 & 228 \\
3877 & 22 54 19 & +67 59.2 & 14.5 & 0.748 &  63.2 &55348 & 22 54 19 &(+68 15.5)& 14.5 & 0.748 &  64 \\
3975 & 23 32 08 & -47 11.5 & 12.4 & 0.531 & 126.2 &57267 & 23 32 06 &(-46 12.0)& 12.4 & 0.530 & 126 \\
\enddata                              
\tablenotetext{a}{Stars with $\mu>0.5\arcsec$ yr$^{-1}$ listed in both
the LHS and NLTT catalogs, but for which one of the coordinates has a
positional error of $5\arcmin$ or larger. Data from the original
tables are given here, with the erroneous entry shown in parenthesis.}
\tablenotetext{b}{LHS catalog designation.}
\tablenotetext{c}{NLTT catalog record number (``recno'' in ViZier).}
\end{deluxetable}                     


\begin{deluxetable}{rrrrrr} 
\tablecolumns{6} 
\tablewidth{0pc} 
\tabletypesize{\scriptsize}
\tablecaption{$\mu>0.5\arcsec$ yr$^{-1}$ LHS stars not listed in the NLTT catalog\tablenotemark{a}}
\tablehead{ 
\colhead{LHS\tablenotemark{b}} & 
\colhead{RA1950} &
\colhead{DE1950} &
\colhead{r} & 
\colhead{pm($\arcsec$ yr$^{-1}$)} & 
\colhead{pma($^{\circ}$)}
}

\startdata 
2397a & 11 19 20 & -12 56.6 & 18.3 &   0.515 & 261.0 \\
325a  & 12 21 24 & -27 41.4 & 17.9 &   1.293 & 284.4 \\
534   & 23 04 36 & +71 26.8 & 11.9 &   1.320 &  71.5 \\
3989  & 23 36 01 & -31 28.0 & 13.0 &   0.701 & 170.3 \\
\enddata
\tablenotetext{a}{All values in this table are quoted from the LHS catalog}
\tablenotetext{b}{LHS catalog designation.}
\end{deluxetable}


\begin{deluxetable}{rrrrrr} 
\tablecolumns{6} 
\tablewidth{0pc} 
\tabletypesize{\scriptsize}
\tablecaption{$\mu>0.5\arcsec$ yr$^{-1}$ NLTT stars not listed in the LHS catalog\tablenotemark{a}}
\tablehead{ 
\colhead{NLTT\tablenotemark{b}} & 
\colhead{RA1950} &
\colhead{DE1950} &
\colhead{r} & 
\colhead{pm($\arcsec$ yr$^{-1}$)} & 
\colhead{pma($^{\circ}$)}
}

\startdata 
 5014 & 01 28 10 & -30 30.6 & 14.6 &    0.639 & 113 \\
11999 &(03 48 27)&(+00 01.3)\tablenotemark{c}& 16.5 &    0.525 & 176\\
13844 & 04 41 29 & +43 19.7 & 16.3 &    0.720 & 152\\
14709 & 05 11 20 & +44 29.0 & 10.0 &    0.660 & 178\\
14710 & 05 11 20 & +44 29.0 & 13.5 &    0.660 & 178\\
22764 &(09 48 37)&(+44 56.3)& 18.1 &    1.049 & 315\\
24136 & 10 19 24 & +44 25.3 & 18.1 &   (2.375)\tablenotemark{d}& 166\\
39081 & 14 59 06 & -05 12.9 & 12.4 &    0.530 & 185\\
44554 & 17 16 08 & -29 46.0 & 14.3 &    0.536 & 188\\
44556 & 17 16 09 & -29 46.1 & 14.5 &    0.536 & 188\\
44557 & 17 16 09 & -29 46.1 & 16.0 &    0.536 & 188\\
44808 & 17 24 41 & -32 38.2 & 18.8 &    0.570 & 240\\
44923 & 17 28 17 & -29 33.8 & 16.1 &    0.500 & 208\\
44952 & 17 28 50 & -29 49.0 & 15.0 &    0.585 & 215\\
46347 & 18 19 13 & +00 16.1 & 15.4 &    0.612 & 201\\
46407 & 18 21 15 & -04 08.2 & 17.6 &    0.608 & 222\\
46583 & 18 28 02 & -13 02.9 & 16.4 &    0.897 & 194\\
46584 &(18 28 02)&(-14 02.6)& 16.6 &    0.812 & 196\\
48293 & 19 47 28 & -29 24.2 & 14.0 &   (3.204)& 187\\
48304 & 19 47 55 & -21 56.2 & 16.9 &   (3.242)& 167\\
48333 & 19 49 17 & -24 33.3 & 14.0 &   (3.259)& 192\\
50541 & 21 04 12 & -73 22.0 &  5.7 &    0.550 & 127\\
50542 & 21 04 12 & -73 22.0 &      &    0.550 & 127\\
54033 & 22 28 07 & +58 45.1 & 15.3 &    0.757 &  66\\
56871 & 23 24 02 & -22 59.3 & 13.8 &    0.598 & 170\\
\enddata
\tablenotetext{a}{All values in this table are quoted from the NLTT catalog}
\tablenotetext{b}{NLTT catalog record number (``recno'' in ViZier).}
\tablenotetext{c}{Stars with coordinates in parenthesis could not be
 found in the DSS within $5\arcmin$ of the quoted position and may not
 exist.}
\tablenotetext{d}{Stars with proper motion in parenthesis were
 recovered in the DSS but found to have actual proper motion 1/10th of
 the quoted value.}
\end{deluxetable}


\begin{deluxetable}{lrrrrrr}
\tabletypesize{\scriptsize}
\tablecolumns{7}
\tablewidth{340pt}
\tablecaption{New stars with $\mu>0.5\arcsec$ yr$^{-1}$ at low galactic
latitudes}
 \tablehead{
\colhead{Star} & 
\colhead{$\alpha$(2000)} &
\colhead{$\delta$(2000)} &
\colhead{pm($\arcsec$ yr$^{-1}$)} & 
\colhead{pma($^{\circ}$)} &
\colhead{$r$} & 
\colhead{$b$}
}
\startdata 
LSR0002+6357 &  0 02 22.58& +63 57 43.6&  0.910&  84.1& 16.3& 17.8\\
LSR0011+5908 &  0 11 31.80& +59 08 39.8&  1.483& 218.3& 14.5& 17.3\\
LSR0013+5437 &  0 13 24.46& +54 37 57.7&  0.981&  52.5& 17.4& 19.4\\
LSR0014+6546 &  0 14 40.15& +65 46 44.8&  0.962&  67.2& 15.7& 18.0\\
LSR0020+5526 &  0 20 08.16& +55 26 32.6&  0.541&  55.7& 14.9& 17.2\\
LSR0031+7732 &  0 31 28.51& +77 32 11.8&  0.600&  88.6& 15.3& 17.4\\
LSR0124+6819 &  1 24 23.38& +68 19 31.1&  0.559& 131.4& 16.4& 18.9\\
LSR0131+5246 &  1 31 51.91& +52 46 31.4&  0.525& 100.8& 18.4& \nodata\\
LSR0134+6459 &  1 34 11.45& +64 59 36.2&  0.924&  75.5& 15.2& 17.9\\
LSR0155+3758 &  1 55 02.33& +37 58 02.3&  0.538& 153.9& 14.8& 17.0\\
LSR0157+5308 &  1 57 40.63& +53 08 13.2&  0.641&  94.7& 14.9& 17.1\\
LSR0200+5530 &  2 00 24.94& +55 30 21.2&  0.507& 112.0& 19.7& \nodata\\
LSR0212+7012 &  2 12 51.67& +70 12 29.9&  0.744&  76.2& 15.9& 18.4\\
LSR0254+3419 &  2 54 44.26& +34 19 01.6&  0.892& 132.5& 19.2& \nodata\\
LSR0258+5354 &  2 58 45.43& +53 54 47.2&  0.545& 127.7& 14.9& 17.2\\
LSR0310+6634 &  3 10 57.65& +66 34 02.6&  0.808& 121.6& 17.5& 19.1\\
LSR0316+3132 &  3 16 22.75& +31 32 44.2&  0.759& 139.2& 15.4& 17.7\\
LSR0340+5124 &  3 40 26.21& +51 24 59.0&  0.932& 155.9& 15.4& 17.6\\
LSR0342+5527 &  3 42 53.74& +55 27 30.2&  0.500& 145.8& 14.8& 17.0\\
LSR0346+2456 &  3 46 46.54& +24 56 02.8&  1.260& 155.2& 18.2& \nodata\\
LSR0354+3333 &  3 54 00.91& +33 33 31.0&  0.848& 150.2& 16.4& 19.2\\
LSR0358+8111 &  3 58 48.55& +81 11 19.0&  0.547& 143.8& 16.9& 19.0\\
LSR0359+4236 &  3 59 09.22& +42 36 33.5&  0.501& 154.0& 15.2& 17.5\\
LSR0400+5417 &  4 00 12.84& +54 17 29.4&  0.754& 136.7& 16.0& 18.0\\
LSR0401+5131 &  4 01 01.51& +51 31 30.0&  0.886& 156.0& 16.7& 17.9\\
LSR0403+2616 &  4 03 07.58& +26 16 11.3&  0.667& 135.7& 19.1& \nodata\\
LSR0419+4233 &  4 19 52.15& +42 33 30.6&  1.535& 159.4& 17.4& \nodata\\
LSR0420+5624 &  4 20 48.05& +56 24 20.2&  0.514& 106.1& 18.3& \nodata\\
LSR0455+0244 &  4 55 30.31& +02 44 23.6&  0.768& 123.2& 16.4& 18.5\\
LSR0455+5252 &  4 55 11.83& +52 52 28.9&  0.804& 187.8& 18.3& \nodata\\
LSR0505+3043 &  5 05 11.76& +30 43 31.8&  1.097& 148.5& 16.0& 18.5\\
LSR0505+6633 &  5 05 40.15& +66 33 51.1&  0.577& 155.1& 17.0& \nodata\\
LSR0510+2712 &  5 10 17.14& +27 12 43.2&  0.652& 131.1& 16.2& 18.2\\
LSR0510+2713 &  5 10 18.31& +27 13 21.4&  0.662& 199.2& 15.4& \nodata\\
LSR0515+5911 &  5 15 30.98& +59 11 17.2&  1.015& 173.3& 16.8& \nodata\\
LSR0519+4213 &  5 19 13.66& +42 13 49.1&  1.181& 119.3& 16.1& 18.8\\
LSR0519+6443 &  5 19 00.50& +64 43 59.2&  0.528& 139.5& 18.0& \nodata\\
LSR0520+2159 &  5 20 53.50& +21 59 44.5&  0.737& 126.9& 16.4& 19.0\\
LSR0521+3425 &  5 21 01.94& +34 25 11.3&  0.512& 149.2& 15.8& 18.1\\
LSR0522+3814 &  5 22 05.33& +38 14 15.4&  1.703& 164.1& 14.5& 16.6\\
LSR0524+3358 &  5 24 41.97& +33 58 23.2&  0.530& 138.4& 17.5& \nodata\\
LSR0527+0019 &  5 27 47.62& -00 19 36.8&  0.530&  95.9& 18.3& \nodata\\
LSR0527+3009 &  5 27 48.05& +30 09 14.8&  0.639& 221.0& 16.3& 18.5\\
LSR0530+5928 &  5 30 39.43& +59 28 23.5&  0.529& 112.0& 18.2& \nodata\\
LSR0532+1354 &  5 32 43.27& +13 54 01.4&  0.578&  47.4& 19.6& \nodata\\
LSR0533+3837 &  5 33 20.40& +38 37 15.2&  0.551& 134.9& 16.3& 18.1\\
LSR0534+2820 &  5 34 11.11& +28 20 54.6&  0.612& 157.4& 17.7& \nodata\\
LSR0539+4038 &  5 39 24.84& +40 38 43.1&  1.057& 141.8& 17.0& \nodata\\
LSR0541+3959 &  5 41 03.89& +39 59 46.3&  0.566& 143.2& 17.1& 18.1\\
LSR0544+2603 &  5 44 57.67& +26 03 00.0&  1.702& 144.3& 16.1& 18.0\\
LSR0549+2329 &  5 49 35.47& +23 29 53.2&  1.379& 131.4& 17.4& 19.1\\
LSR0550+8130 &  5 50 05.83& +81 30 46.1&  0.588& 145.8& 16.2& 18.9\\
LSR0556+1144 &  5 56 57.26& +11 44 33.7&  0.611& 121.8& 14.2& 16.5\\
LSR0557+0036 &  5 57 00.15& -00 36 24.5&  0.545&  90.3& 15.8& \nodata\\
LSR0557+0435 &  5 57 35.11& +04 35 12.8&  0.506& 140.3& 17.1& 18.0\\
LSR0602+3910 &  6 02 30.48& +39 10 58.8&  0.522& 163.8& 18.3& \nodata\\
LSR0606+1706 &  6 06 54.15& +17 06 13.3&  0.517& 145.5& 18.1& \nodata\\
LSR0609+2319 &  6 09 52.44& +23 19 12.7&  1.104& 130.7& 16.5& 18.7\\
LSR0612+1002 &  6 12 53.95& +10 02 32.6&  0.599& 285.3& 19.8& \nodata\\
LSR0618+1614 &  6 18 52.54& +16 14 56.0&  0.646& 165.7& 14.7& 16.9\\
LSR0621+1219 &  6 21 34.51& +12 19 43.7&  0.513& 148.9& 17.3& \nodata\\
LSR0621+3652 &  6 21 31.61& +36 52 57.4&  0.865& 176.9& 14.5& 16.8\\
LSR0627+0616 &  6 27 33.31& +06 16 58.8&  1.019& 178.5& 15.4& 17.6\\
LSR0628+0529 &  6 28 43.44& +05 29 13.9&  0.551& 140.4& 17.1& \nodata\\
LSR0638+3128 &  6 38 00.84& +31 28 53.8&  0.614& 156.1& 19.5& \nodata\\
LSR0639+2920 &  6 39 05.50& +29 20 33.7&  0.500& 173.6& 17.2& \nodata\\
LSR0646+3212 &  6 46 15.00& +32 12 01.4&  0.567& 142.4& 17.5& \nodata\\
LSR0658+4442 &  6 58 54.38& +44 42 16.2&  0.685& 173.0& 18.8& \nodata\\
LSR0702+2154 &  7 02 30.53& +21 54 30.2&  0.653& 144.7& 14.9& 16.9\\
LSR0705+0506 &  7 05 48.77& +05 06 17.3&  0.510& 158.4& 16.0& 18.8\\
LSR0721+3714 &  7 21 25.39& +37 14 05.6&  0.587& 234.7& 16.2& 18.7\\
LSR0723+3806 &  7 23 49.99& +38 06 02.5&  0.831&  92.7& 19.1& \nodata\\
LSR0731+0729 &  7 31 29.76& +07 29 59.3&  0.512& 190.3& 15.2& 17.4\\
LSR0745+2627 &  7 45 08.98& +26 27 06.5&  0.884& 144.0& 18.4& \nodata\\
LSR0803+1548 &  8 03 30.00& +15 48 09.4&  0.516& 154.3& 16.1& 18.4\\
LSR1722+1004 & 17 22 32.54& +10 04 06.2&  0.724& 192.1& 15.1& 17.5\\
LSR1755+1648 & 17 55 32.76& +16 48 59.0&  0.995& 116.9& 14.2& 16.4\\
LSR1757+0015 & 17 57 51.00& -00 15 08.3&  0.518& 184.8& 17.0& \nodata\\
LSR1758+1417 & 17 58 22.92& +14 17 38.0&  1.014& 235.4& 15.8& 16.9\\
LSR1802+0028 & 18 02 19.08& -00 28 40.8&  0.543& 211.3& 17.6& \nodata\\
LSR1806+1141 & 18 06 21.69& +11 41 00.2&  0.541& 186.0& 14.4& 16.5\\
LSR1808+1134 & 18 08 12.15& +11 34 46.6&  0.606& 228.7& 14.4& 16.8\\
LSR1809-0219 & 18 09 43.66& -02 19 35.0&  0.506& 176.0& 13.9& 16.4\\
LSR1809-0247 & 18 09 50.14& -02 47 43.1&  1.005& 214.9& 15.2& 17.4\\
LSR1817+1328 & 18 17 06.48& +13 28 25.0&  1.207& 201.5& 15.5& 16.7\\
LSR1820-0031 & 18 20 49.66& -00 31 26.0&  0.555& 199.3& 15.7& 17.8\\
LSR1833+2219 & 18 33 14.76& +22 19 17.4&  0.502& 200.0& 14.9& 17.6\\
LSR1835+3259 & 18 35 37.90& +32 59 53.5&  0.747& 185.8& 16.6& \nodata\\
LSR1836+1040 & 18 36 03.57& +10 40 07.3&  0.921& 206.0& 15.9& 18.2\\
LSR1841+2421 & 18 41 47.90& +24 21 59.8&  0.752& 189.0& 16.5& 19.0\\
LSR1843+0507 & 18 43 52.63& +05 07 13.4&  0.577& 253.2& 17.1& 19.2\\
LSR1844+0947 & 18 44 38.93& +09 47 57.5&  0.501& 224.3& 15.8& 17.6\\
LSR1851+2641 & 18 51 45.24& +26 41 57.1&  0.704&  26.3& 16.1& 18.9\\
LSR1859+0156 & 18 59 28.01& +01 56 02.4&  0.674& 142.2& 14.7& 16.9\\
LSR1914+2825A& 19 14 05.54& +28 25 52.3&  0.529& 212.6& 15.0& 17.6\\
LSR1914+2825B& 19 14 06.02& +28 25 45.5&  0.531& 214.3& 15.6& 18.7\\
LSR1915+1609 & 19 15 40.94& +16 09 29.9&  1.572& 161.2& 17.4& \nodata\\
LSR1918+1728 & 19 18 36.99& +17 28 00.1&  0.626& 192.2& 17.2& \nodata\\
LSR1919+1438 & 19 19 27.84& +14 38 02.4&  0.507&  96.6& 14.9& 16.7\\
LSR1922+4605 & 19 22 57.26& +46 05 14.3&  0.555& 206.3& 15.1& 17.8\\
LSR1927+6802 & 19 27 09.55& +68 02 22.9&  0.515& 227.2& 17.2& 19.4\\
LSR1928-0200A& 19 28 13.20& -02 00 17.3&  0.858& 194.7& 15.2& 17.2\\
LSR1928-0200B& 19 28 13.35& -02 00 07.5&  0.858& 194.7& 18.2& \nodata\\
LSR1933-0138 & 19 33 59.50& -01 38 18.2&  0.895& 132.6& 13.5& 15.5\\
LSR1943+0941 & 19 43 13.56& +09 41 24.4&  0.543& 213.6& 16.2& 18.7\\
LSR1945+4650A& 19 45 30.36& +46 50 15.7&  0.612& 228.0& 16.8& 17.9\\
LSR1945+4650B& 19 45 30.36& +46 50 06.7&  0.609& 228.7& 17.0& 18.1\\
LSR1946+0937 & 19 46 29.83& +09 37 12.4&  0.566& 173.7& 16.8& 17.5\\
LSR1946+0942 & 19 46 17.52& +09 42 47.9&  0.555& 173.8& 14.6& 17.0\\
LSR1956+4428 & 19 56 17.48& +44 28 46.9&  0.891& 214.3& 15.0& 17.3\\
LSR2000+0404 & 20 00 05.20& +04 04 42.2&  0.503& 125.1& 15.7& 18.0\\
LSR2000+3057 & 20 00 05.71& +30 57 32.0&  1.339&  16.7& 15.6& 17.7\\
LSR2005+0835 & 20 05 01.77& +08 35 15.0&  0.583& 197.9& 14.6& 16.5\\
LSR2009+5659 & 20 09 33.82& +56 59 25.8&  0.824&  31.5& 14.2& 16.3\\
LSR2010+3938 & 20 10 53.31& +39 38 08.5&  0.512& 180.4& 13.3& 17.4\\
LSR2013+0417 & 20 13 11.98& +04 17 20.4&  0.749& 195.5& 14.4& 16.2\\
LSR2017+0623 & 20 17 32.50& +06 23 51.7&  0.674& 187.3& 16.0& 18.2\\
LSR2036+5059 & 20 36 21.86& +50 59 50.3&  1.054& 100.8& 16.8& \nodata\\
LSR2044+1339 & 20 44 22.08& +13 39 01.1&  0.518&  31.2& 15.4& 17.5\\
LSR2050+7740 & 20 50 05.59& +77 40 24.6&  0.543&  27.3& 18.0& 19.3\\
LSR2059+5517 & 20 59 44.95& +55 17 30.5&  0.500&  33.1& 16.8& 18.2\\
LSR2105+2514 & 21 05 16.58& +25 14 48.1&  0.563& 150.2& 16.1& 18.7\\
LSR2107+3600 & 21 07 47.50& +36 00 00.7&  0.736& 231.4& 15.9& 18.4\\
LSR2115+3804 & 21 15 31.63& +38 04 39.7&  0.506&  25.2& 15.0& 17.3\\
LSR2117+7345 & 21 17 41.73& +73 45 54.7&  0.746&  41.1& 16.4& 18.6\\
LSR2122+3656 & 21 22 56.35& +36 56 00.2&  0.816&  65.0& 16.2& 18.7\\
LSR2124+4003 & 21 24 32.33& +40 03 59.4&  0.697&  50.9& 14.9& 17.3\\
LSR2132+4754 & 21 32 05.61& +47 54 41.0&  0.569&  37.8& 14.5& 16.8\\
LSR2146+5147 & 21 46 34.56& +51 47 33.7&  0.584&  49.7& 16.0& 18.5\\
LSR2157+2705 & 21 57 59.57& +27 05 19.0&  0.502&  45.6& 17.5& 18.9\\
LSR2158+6117 & 21 58 34.63& +61 17 06.0&  0.819&  82.3& 15.7& 18.1\\
LSR2205+5353 & 22 05 47.09& +53 53 44.9&  0.528& 236.4& 15.1& 17.3\\
LSR2205+5807 & 22 05 32.78& +58 07 26.8&  0.538& 108.9& 17.6& \nodata\\
LSR2214+5211 & 22 14 05.28& +52 11 36.2&  0.574& 139.0& 16.8& 18.6\\
LSR2251+4706 & 22 51 52.68& +47 06 13.0&  0.631&  71.5& 17.9& \nodata\\
LSR2311+5032 & 23 11 53.64& +50 32 15.0&  0.669&  81.0& 14.6& 16.9\\
LSR2311+5103 & 23 11 15.65& +51 03 58.0&  0.531& 253.5& 18.1& \nodata\\
LSR2321+4704 & 23 21 23.31& +47 04 38.6&  0.712&  70.5& 16.2& 18.5\\
LSR2345+4942 & 23 45 40.80& +49 42 29.5&  0.576& 231.7& 13.0& 15.2\\
LSR2356+5311 & 23 56 29.59& +53 11 01.7&  0.635& 118.0& 14.5& 16.9\\
LSR2357+5528 & 23 57 44.64& +55 28 16.3&  0.507&  85.7& 14.0& 16.5\\
\enddata                              
\end{deluxetable}                     

\begin{figure}
\caption{How the SUPERBLINK software works. A pair of POSS-I and
POSS-II images of a region of the sky (top: here a crowded field in Cygnus)
are given as input. The code first remaps the two images to a common
grid, and renormalizes the intensity scale (center). The images are
then shifted and rotated to minimize the residuals, and the best image
(usually POSS-II) is degraded to match the quality of the other
(bottom). The final residuals map is then used to identify moving
objects in the fields, which show up as pairs of high intensity
features.}
\end{figure}

\begin{figure}
\caption{Recovery rate by SUPERBLINK of high proper motion stars with
$\mu>0.5\arcsec$ yr$^{-1}$ listed in the Tycho and Tycho-2 catalog in
the area of the survey. Our DSS-based survey did recover all the stars
in the last magnitude bin (which contains 15 stars).}
\end{figure}

\begin{figure}
\caption{Difference between the quoted positions of Luyten stars
recovered in our survey area and our own measured positions. The
majority of the stars were recovered within 10$\arcsec$ of the
best position quoted by Luyten (from either the LHS or NLTT). All the
stars were recovered within 50$\arcsec$ of the quoted positions except
LHS6374 (not shown of the plots above), which we recovered
563$\arcsec$ from the best position quoted by Luyten.}
\end{figure}

\begin{figure}
\caption{Distribution of high proper motion stars in the area covered
by our survey. The left panel is a histogram of the stars found as a
function of the POSS-II red magnitude (for the brightest stars with no
POSS-II magnitudes available, the Luyten-r magnitude is used instead). 
The right panels show the distribution of stars in galactic
coordinates. Black refers to stars whose proper motion has been
measured by the HIPPARCOS satellite and found in the Hipparcos and
Tycho catalog, or in the Tycho-2 catalog. Blue refers to high proper
motion stars not observed by the HIPPARCOS but listed in the Luyten
catalogs (LHS and/or NLTT) and recovered in the DSS with our
SUPERBLINK software. Red refers to stars listed in the Luyten catalogs
that were not observed with HIPPARCOS and that were not recovered with
SUPERBLINK. Green refers to new high proper motion stars discovered in
the DSS with SUPERBLINK.}
\end{figure}

\begin{figure}
\caption{Estimated completeness in high proper motion stars as a
function of POSS-II red magnitude for the surveys discussed in
this paper. This is for the low-galactic latitude ($|b|<25^{\circ}$)
survey area only. The completeness of the Luyten and Tycho catalogs
are estimated based on the number of high proper motion stars
currently known (this paper). The completeness of the sample of stars
recovered in the DSS by the SUPERBLINK software is based on the
recovery of Luyten stars. The Luyten catalogs are less than $50\%$
complete beyond $r=15$, while our DSS survey is more than $95\%$
complete for faint stars down to $r=19$.}
\end{figure}

\begin{figure}
\caption{Comparison between magnitudes of high proper motion stars
quoted in the Tycho-2 catalog ($V_T$, $B_T$), in the Luyten catalogs
(Luyten-b, Luyten-r), and in the GSC2.2.1 catalog (POSS-II magnitudes
$r$ and $b$). Using the Tycho-2 magnitudes as a standard, we find that
the POSS-II magnitudes from the GSC2.2.1 are significantly more
accurate that the Luyten magnitudes. The Luyten ``red'' magnitude
(Luyten-r) is also found to be more representative of a visual
magnitude, as it matches $V_T$ relatively well.}
\end{figure}

\begin{figure}
\caption{Comparison between the colors of high proper motion stars
quoted in different catalogs, as in Figure 6. Comparison between the
colors of high proper motion stars found in both the Tycho-2 and
Luyten catalogs shows that the Luyten blue-red color is actually more
consistent with a blue-visual color. The POSS-II $b-r$ color index has
systematically larger values than the Tycho and Luyten colors indexes,
as it is effectively a blue-red color. The large scatter in the colors
of HPM stars from the POSS-II and from Luyten's catalogs suggests that
many HPM stars for which only the Luyten magnitudes are available may
have been misclassified.}
\end{figure}

\begin{figure}
\caption{Comparison of reduced proper motion diagrams built from the
Luyten catalog (with the Luyten-r and Luyten-b magnitudes) and from
our updated catalog (with POSS-II magnitudes from the GSC2.2.1), for
stars within the area of our survey. Left panel: reduced proper motion
diagrams for the subsample of Luyten stars for which we also have
POSS-II colors, with the color and reduced proper motion calculated
from the Luyten-r and Luyten-b magnitudes. Middle panel: same stars as
in the left panel, but now with color and reduced proper motion
calculated with the POSS-II $b$ and $r$ magnitudes. Notice the smaller
dispersion in color and the better separation between the red dwarf
and white dwarf sequences. Right panel: reduced proper motion diagram
for all the stars in our catalog for which we have POSS-II $r$ and $b$
magnitudes. The new census nicely extends both the red dwarf and white
dwarf sequences to larger reduced proper motions.}
\end{figure}

\begin{figure}
\caption{Distribution of high proper motion stars ($\mu>0.5\arcsec$
yr$^{-1}$ as a function of the galactic latitude $b$ for the survey
area. The black-shaded region shows the distribution of Luyten stars
with red magnitudes in the range $13<r<16$ (left panel) and $16<r<20$
(right panel). The white-shaded region represents the new stars
discovered with SUPERBLINK and discussed in this paper. The Luyten
survey was very deficient in fainter stars, even at relatively high
($|b|\sim20$) galactic latitudes. The updated census is uniform over
$b$, which suggests that the survey for high proper motion stars is
now complete at low galactic latitudes.}
\end{figure}

\begin{figure}
\caption{Finding charts for the new high proper motion stars
discovered in our survey.}
\end{figure}

\begin{figure}
\caption{Finding charts for the new high proper motion stars (continued).}
\end{figure}

\begin{figure}
\caption{Finding charts for the new high proper motion stars (continued).}
\end{figure}

\begin{figure}
\caption{Finding charts for the new high proper motion stars (continued).}
\end{figure}

\begin{figure}
\caption{Finding charts for the new high proper motion stars (continued).}
\end{figure}

\begin{figure}
\caption{Finding charts for the new high proper motion stars (continued).}
\end{figure}

\begin{figure}
\caption{Finding charts for the new high proper motion stars (continued).}
\end{figure}

\end{document}